\documentclass[aps,prl,twocolumn,showpacs,groupedaddress]{revtex4-1}
\usepackage{epsfig}
\usepackage{longtable}
\usepackage{amsmath,amssymb,amsfonts,amsthm,bbm}
\usepackage[hang,nooneline]{subfigure}
\usepackage{enumerate}
\usepackage{pstricks}

\newcommand{\be}{\begin{equation}}
\newcommand{\ee}{\end{equation}}

\newcommand{\vek}[2]{ \left( \begin{smallmatrix} #1\\#2 \end{smallmatrix} \right) }

\newcommand{\comment}[1]{}

\begin{document}


\title{Complete analytic solution of the geodesic equation \\ in Schwarzschild--(anti) de Sitter space--times}

\author{Eva Hackmann  and Claus L{\"a}mmerzahl}
\address{ZARM, University of Bremen, Am Fallturm, 28359 Bremen, Germany}

\date{April 17, 2008}

\begin{abstract}
The complete set of analytic solutions of the geodesic equation in a Schwarzschild--(anti) de Sitter space--time is presented. The solutions are derived from the Jacobi inversion problem restricted to the theta--divisor. In its final form the solutions can be expressed in terms of derivatives of Kleinian sigma functions. The solutions are completely classified by the structure of the zeros of the characteristic polynomial which depends on the energy, angular momentum, and the cosmological constant. 
\end{abstract}


\maketitle

\paragraph{Introduction and motivation}

The Einstein equations with a cosmological constant $R_{\mu\nu} - \frac{1}{2} R + \Lambda g_{\mu\nu} = \kappa T_{\mu\nu}$ today are taken as very successful model to incorporate the observation of an accelerated expanding universe \cite{PeeblesRatra03}. Within this theory the most simple solution of an isolated mass is the spherically symmetric Schwarzschild--de Sitter solution (e.g. \cite{Rindler01})
\begin{equation}\label{metric}
ds^2 = \alpha dt^2 - \alpha^{-1} dr^2 - r^2 (d\theta^2 + \sin^2\theta d\varphi) \, ,
\end{equation}
where $\alpha = 1 - \frac{r_{\rm S}}{r} - \frac{1}{3} \Lambda r^2$ (we take $c = G = 1$). This Schwarzschild--de Sitter metric is characterized by the Schwarzschild--radius $r_{\rm S} = 2 M$ related to the mass $M$ of the gravitating body, and the cosmological constant $\Lambda$ which today is assumed to have the value $\sim 10^{-52}\;{\rm m}^{-2}$ \cite{PeeblesRatra03}. For a general discussion of this metric, see e.g. \cite{Rindler01,Geyer80}. This may serve as a model for the Solar system in an expanding universe. Recently such a model has attracted some interest in order to discuss the possibility to observe the influence of $\Lambda$ on moving masses on the scale of the Solar system \cite{KerrHauckMashhoon03}. This requires a discussion of the geodesic equation and its solutions in such a Schwarzschild--de Sitter space-time. Exact solutions of the geodesic equation are also of interest in the calculation of gravitational wave templates for Extreme Mass Ratio Inspirals (EMRI) \cite{BarackCutler07}. There is further interest to understand explicitly the structure of geodesics in the background of black holes in anti-de Sitter space in the context of string theory and the AdS/CFT correspondence. In addition, recently there also has been a lot of work dealing with geodesics and integrability in black hole backgrounds in higher dimensions, also in the presence of a cosmological constant \cite{FrolovStojkovic03}.

Here we present explicitly the full analytic solution of the geodesic equation in a Schwarzschild--(anti)de Sitter space--time and their complete classification. So far, explicit solutions for the geodesic equations are only known for the Schwarzschild solution \cite{Hagihara31}, the Reissner--Nordstr\"om solution \cite{Chandrasekhar83}, and for the Kerr and Kerr--Newman space--time \cite{Chandrasekhar83}. For a specialized case orbits in a Schwarzschild--(anti) de Sitter space--time have been presented in  \cite{CruzOlivaresVillanueva05}.

\paragraph{The geodesic equation}

We consider the geodesic equation
\begin{equation}
0 = \frac{d^2 x^\mu}{ds^2} + \left\{\begin{smallmatrix} \mu \\ \rho\sigma \end{smallmatrix}\right\} \frac{dx^\rho}{ds} \frac{dx^\sigma}{ds} \, ,
\end{equation}
where $ds^2 = g_{\mu\nu} dx^\mu dx^\nu$ is the proper time and $\left\{\begin{smallmatrix} \mu \\ \rho\sigma \end{smallmatrix}\right\}$ is the Christoffel symbol in a space--time given by the metric \eqref{metric}. The geodesic equation has to be supplemented by the normalization condition $g_{\mu\nu} \frac{dx^\mu}{ds} \frac{dx^\nu}{ds} = \epsilon$ where for massive particles $\epsilon = 1$ and for light $\epsilon = 0$.   

Due to the spherical symmetry we can restrict our consideration to the equatorial plane. Furthermore, due to the conserved energy $E$ and angular momentum $L$ the geodesic equation reduces to one ordinary differential equation \comment{Aenderung!!}
\begin{equation}
\frac{1}{2} \left( \frac{dr}{ds} \right)^2 = E - V_{\rm eff}(r)\,,
\end{equation}
with the effective potential
\begin{equation}\label{potential}
V_{\rm eff}(r) = \frac{1}{2} \left( -\frac{1}{3} \Lambda L^2 - \epsilon \frac{r_S}{r} + \frac{L^2}{r^2} - \frac{r_SL^2}{r^3} - \frac{\epsilon}{3}\Lambda r^2 \right)\,.
\end{equation}
A substitution $u = r_{\rm S}/r$ gives
\begin{equation}
\left( \frac{du}{d\varphi} \right)^2 = u^3 - u^2 + \epsilon \lambda u + \lambda(\mu-\epsilon) + \frac{1}{3} \rho + \frac{\epsilon}{3}  \lambda \rho \frac{1}{u^2}  \label{Dgl3}
\end{equation}
with the dimensionless parameters
\begin{equation}\label{para}
\lambda := \frac{r_{\rm S}^2}{L^2} \geq 0 \,, \quad \mu := E^2 \geq 0 \quad \text{and} \quad \rho:=\Lambda r_{\rm S}^2 \, .
\end{equation}
We rewrite \eqref{Dgl3} as
\begin{equation}\label{Dgl}
\left( u \frac{du}{d\varphi} \right)^2 = P_5(u)
\end{equation}
with
\begin{equation}
P_5(u) := u^5 - u^4 + \epsilon \lambda u^3 + \left( \lambda(\mu-\epsilon) + \frac{1}{3} \rho \right) u^2 + \frac{\epsilon}{3} \lambda \rho \, . \label{P5}
\end{equation}

For light $\epsilon = 0$ the situation simplifies considerably. In this case the solutions can be given in terms of the Weierstrass $\wp$ function \cite{GradshteynRyzhik83}
\begin{equation}
r(\varphi) = \frac{r_S}{4 \wp(\varphi; g_2,g_3) + \frac{1}{3}}\,,
\end{equation}
with the Weierstrass invariants
\begin{equation}
g_2 := \frac{1}{12} \, , \qquad g_3:= - \frac{1}{8} \left(\frac{1}{27} + \frac{1}{2} \left(u_0^3 - u_0^2\right)\right) \, ,
\end{equation}
where $u_0 = r_{\rm S}/b$ and $b$ is the distance of closest approach.

Now we discuss $\epsilon = 1$. Separation of variables in $\eqref{Dgl}$ yields the hyperelliptic integral
\begin{equation}\label{int}
\varphi - \varphi_0 = \int_{u_0}^u \frac{u' du'}{\sqrt{P_5(u')}}\,,
\end{equation}
where $u_0 = u(\varphi_0)$. We explicitly solve this integral by using a method based on the inversion problem of hyperelliptic Abelian integrals \cite{Abel1828}. We use this general ansatz  \cite{KraniotisWhitehouse03} in order to obtain an explicit one--parameter solution of the two--parameter inversion problem by restricting the problem to the so--called theta--divisor. This procedure has been applied previously to the problem of the double pendulum \cite{EnolskiiPronineRichter03}. The resulting orbits can be completely classified. 

In solving \eqref{int} there are two major issues which have to be addressed. First, owing to the two branches of the square root the integrand is not well defined in the complex plane. Second, the solution $u(\varphi)$ should not depend on the integration path. That means that if $\gamma$ denotes some closed integration path and 
\begin{equation}
\oint_\gamma \dfrac{u du}{\sqrt{P_5(u)}} = \omega \label{DefPeriod}
\end{equation}
then
\begin{equation}
\varphi - \varphi_0 - \omega = \int_{u_0}^u \frac{u' du'}{\sqrt{P_5(u')}}
\end{equation}
should be valid, too. Therefore $u$ has to be a periodic functions with period $\omega$. 

\paragraph{The analytic solution}

These problems can be solved if \eqref{int} is considered on the Riemannian surface of $y^2=P_5(x)$. For the construction of periodic functions on this Riemannian surface we introduce the canonical basis of the space of holomorphic differentials $dz_i$ and of associated meromorphic differentials $dr_i$ by
\begin{equation}
\begin{aligned}
dz_1 & := \frac{dx}{\sqrt{P_5(x)}}\,, & dz_2 &:= \frac{x dx}{\sqrt{P_5(x)}}\,, \\
dr_1 &:= \frac{3x^3-2x^2+\lambda x}{4 \sqrt{P_5(x)}} dx \,, & dr_2 &:= \frac{x^2 dx}{4 \sqrt{P_5(x)}}  \,,
\end{aligned}
\end{equation}
as well as the period matrices $( 2\omega, 2\omega^\prime )$ and  $( 2\eta, 2\eta^\prime )$ 
\begin{equation}\label{periodmatrices}
\begin{aligned}
2 \omega_{ij} &:= \oint_{a_j} dz_i\,, & \qquad 2 \omega'_{ij} &:= \oint_{b_j} dz_i\,, \\
2 \eta_{ij} &:= - \oint_{a_j} dr_i\,, & \qquad 2 \eta'_{ij} &:= - \oint_{b_j} dr_i\, .
\end{aligned}
\end{equation}
Here $\{ a_i,b_i \,|\, i=1,2 \} \in H_1(X,\mathbb{Z})$ is the homology basis of closed paths where integration does not evaluate to zero. Finally we introduce the normalized holomorphic differentials 
\begin{equation}
d\vec{v} := (2 \omega)^{-1} d\vec{z} \, , \qquad d\vec z = \begin{pmatrix} dz_1 \\ dz_2 \end{pmatrix} \, .
\end{equation}
The period matrix of these differentials is given by $(\mathbbm{1}_2,\tau)$, where $\mathbbm{1}_2$ is the $2 \times 2$ unit matrix and $\tau$ is defined by
\begin{equation}
\tau:=  \omega^{-1} \omega'\,. \label{normalizedtau}
\end{equation}
It can be shown \cite{Mumford83} that this normalized matrix $\tau$ always is a Riemannian matrix, that is, $\tau$ is symmetric and $\Im \tau$ is positive definite.

The Riemannian surface of $y^2=P_5(x)$ has $4$ independent closed paths, each corresponding to a period of the functions defined on these surfaces and, hence, to a period of the solution $u$ of \eqref{int}. In order to construct functions with four periods, we need the theta function $\vartheta : \mathbb{C}^2 \to \mathbb{C}$, 
\begin{equation}
\vartheta(\vec z;\tau) := \sum_{\vec m \in {\mathbb{Z}}^2} e^{i \pi \vec m^t (\tau \vec m + 2 \vec z)}\, ,
\end{equation}
which is a holomorphic function on $\mathbbm{C}^2$. The theta function is already periodic with respect to the columns of $\mathbbm{1}_2$ and quasi--periodic with respect to the columns of $\tau$, i.e. for any $n \in \mathbb{Z}^2$ it holds
\begin{align}
\vartheta(\vec z + \mathbbm{1}_2 \vec n;\tau) & = \vartheta(\vec z;\tau)\,,\\
\vartheta(\vec z + \tau \vec n;\tau) & = e^{- i \pi \vec n^t (\tau \vec n + 2 \vec z)} \vartheta(\vec z;\tau)\,.
\end{align} 

The solution of \eqref{int} can be explicitly formulated in terms of sigma functions $\sigma : {\mathbb{C}}^2 \rightarrow \mathbb{C}$,
\begin{equation}\label{sigma}
\sigma(\vec z) = C e^{- \frac{1}{2} \vec z^t \eta \omega^{-1} \vec z} \vartheta \left( (2 \omega)^{-1} \vec z + \tau \vek{\frac{1}{2}}{\frac{1}{2}} + \vek{0}{\frac{1}{2}}; \tau \right)\,,
\end{equation}
where the constant $C$ can be given explicitly \cite{BuchstaberEnolskiiLeykin97} but does not matter here. Based on this $\sigma$ are the generalized Weierstrass functions \begin{equation}\label{Weier}
\wp_{ij}(\vec z) = - \frac{\partial}{\partial z_i}  \frac{\partial}{\partial z_j} \log \sigma(\vec z) = \frac{\sigma_i(\vec z) \sigma_j(\vec z) - \sigma(\vec z) \sigma_{ij}(\vec z)}{\sigma^2(\vec z)} \,,
\end{equation}
where $\sigma_{i}$ denotes the derivative of the sigma function with respect to the $i-$th component.

\begin{figure*}[t!]
\subfigure[][$\Lambda = -10^{-45} \text{km}^{-2}$]{
\includegraphics[width=0.2\textwidth]{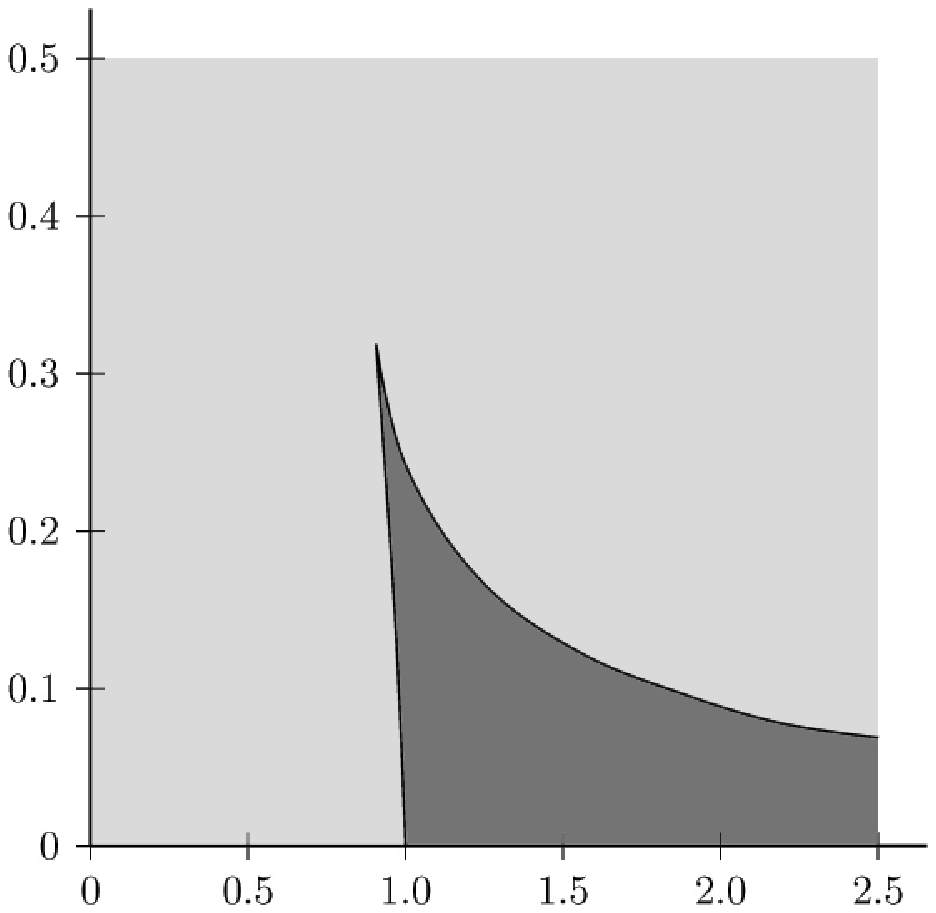}
}
\subfigure[][$\Lambda = 0$]{
\includegraphics[width=0.2\textwidth]{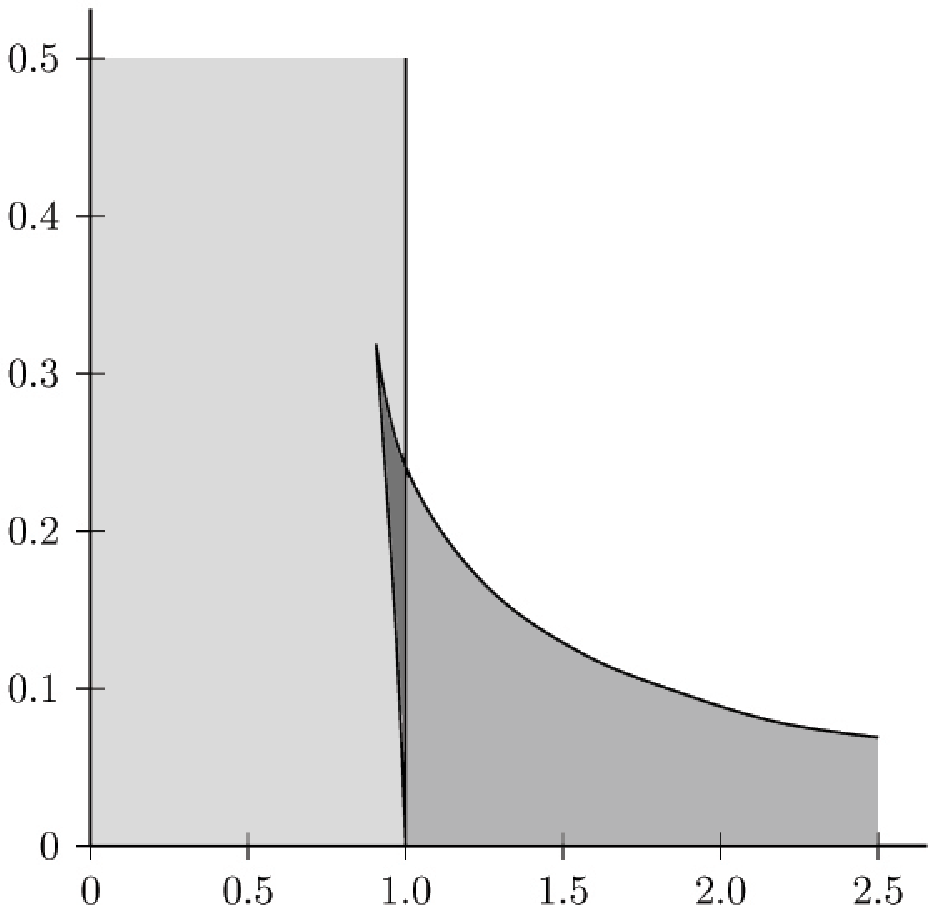}
}
\subfigure[][$\Lambda = 10^{-45} \text{km}^{-2}$]{
\includegraphics[width=0.2\textwidth]{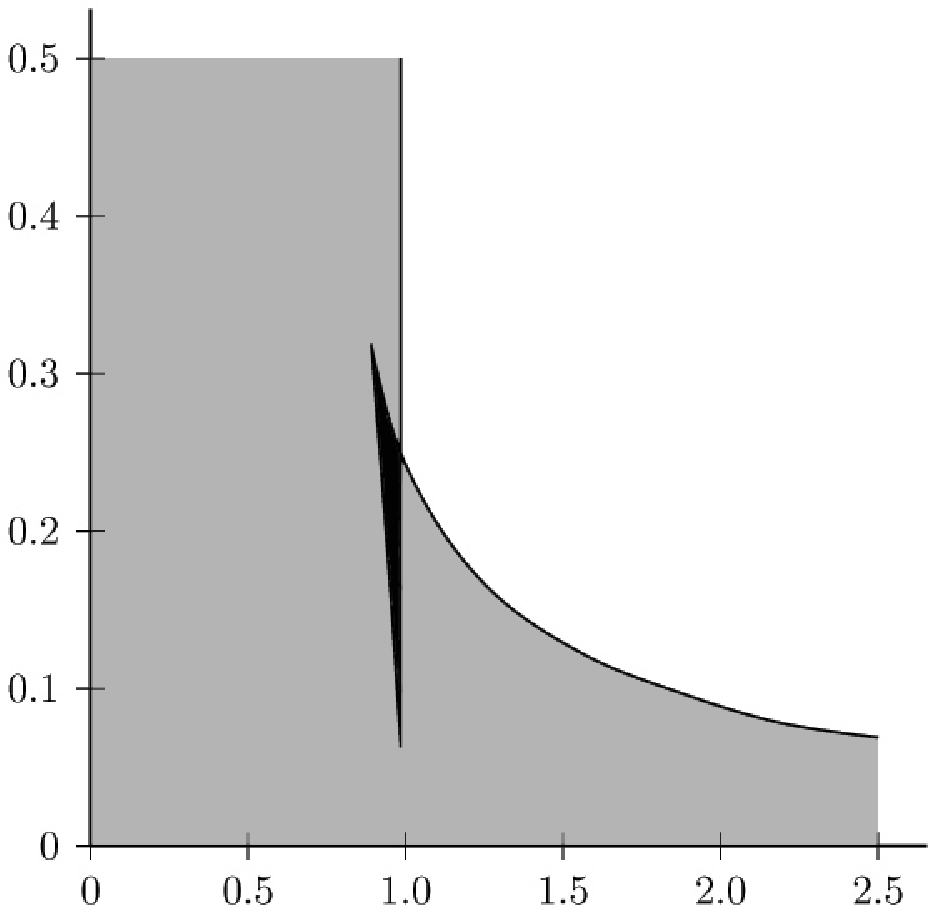}
}
\caption{The zeros of $P_5(u)$ in a $(\mu,\lambda)$--diagram for different values for the cosmological constant ($\mu$ is along the $x$--axis, $\lambda$ along the $y$--axis). The grayscales encode the numbers of positive real zeros of the polynomial $P_5$: black = 4, dark gray = 3, gray = 2, light gray = 1, white = 0. For $\Lambda = 0$ these zeros have been extensively discussed in \cite{Hagihara31}.}
\label{Dia}
\end{figure*}

We need the $2$-dimensional Abel map
\begin{equation}\label{Abel}
\vec A_{x_0} : S^2X \to {\rm Jac}(X)\,, \quad   \begin{pmatrix} x_1 \\ x_2 \end{pmatrix} \mapsto \int_{x_0}^{x_1} d\vec{z} + \int_{x_0}^{x_2} d\vec{z}
\end{equation}
from the $2$nd symmetric power of the Riemannian surface $X$ to the Jacobian ${\rm Jac}(X) = \mathbb{C}^2/\Gamma$ of $X$, where $\Gamma=\{ \omega v + \omega' v' | v, v' \in \mathbb{Z}^2 \}$ is the lattice of periods of the differential $d\vec{z}$. Jacobi's inversion problem is now to determine $\vec x$ for given $\vec \varphi$ from the equation $\vec \varphi = \vec A_{x_0}(\vec x)$, i.e.
\begin{align}\label{Jacobi}
\varphi_1  & = \int_{x_0}^{x_1} \frac{dz}{\sqrt{P_5(z)}} + \int_{x_0}^{x_2} \frac{dz}{\sqrt{P_5(z)}} \,, \nonumber \\ 
\varphi_2  & = \int_{x_0}^{x_1} \frac{z dz}{\sqrt{P_5(z)}} + \int_{x_0}^{x_2} \frac{z dz}{\sqrt{P_5(z)}} \,.
\end{align}
We want to restrict now Jacobi's inversion problem to the set of zeros of $\vartheta(\cdot + \vec K_{x_0}; \tau)$, where $\vec K_{x_0}$ is the vector of Riemann constants, see \cite{BuchstaberEnolskiiLeykin97}. This set is called a theta--divisor and denoted by $\Theta_{\vec K_{x_0}}$. This restriction is necessary since in our case $\vec A^{-1}_{x_0}=\vec \varphi$ should be the (one-dimensional) physical angle and the theta--divisor is a one--dimensional subset of $S^2X$.  

The solution of Jacobi's inversion problem \eqref{Jacobi} can be given in terms of generalized Weierstrass functions. The components of the solution vector $\vec{x}=(x_1,x_2)^t$ are 
\begin{equation}\label{Jacobisol}
\begin{split}
x_1+x_2 & = 4 \wp_{22}(\vec \varphi)\,, \\
x_1 \, x_2 & = - 4 \wp_{12}(\vec \varphi) \, .
\end{split}
\end{equation}
Now we reformulate \eqref{Jacobi} to $\vec \phi=\vec A_{\infty}(\vec x)$ with $\vec \phi = \vec \varphi - \int_{x_0}^\infty d\vec z$ and restrict the problem to the theta--divisor by the limiting process
\begin{align}
x_1 & = \lim_{x_2 \to \infty} \frac{x_1 x_2}{x_1+ x_2} \nonumber\\
& =  \frac{\sigma(\vec\varphi_\infty) \sigma_{12}(\vec\varphi_\infty) - \sigma_1(\vec\varphi_\infty) \sigma_2(\vec\varphi_\infty)}{\sigma_2^2(\vec\varphi_\infty) - \sigma(\vec\varphi_\infty) \sigma_{22}(\vec\varphi_\infty)}
\end{align}
where
\begin{equation}
\vec\varphi_\infty = \lim_{x_2 \to \infty} \vec \varphi = \vec A_\infty(\vec x_\infty) \, , \;\; \vec x_\infty = (x_1,\infty)^t \,. \label{x_1}
\end{equation}
It can be shown, see \cite{Mumford83}, that $(2\omega)^{-1} \vec A_\infty(\vec x_\infty)$ is an element of the theta--divisor $\Theta_{\vec K_\infty}$, where $\vec K_\infty=\tau \vek{\frac{1}{2}}{\frac{1}{2}} + \vek{0}{\frac{1}{2}}$, \cite{BuchstaberEnolskiiLeykin97}. Thus, $\sigma( \vec \varphi_\infty ) =0$ and there is a functional relation between $\varphi_{\infty,1}$ and $\varphi_{\infty,2}$. The solution to the geodesic equation in Schwarzschild--de Sitter space--time is then given by
\begin{align}
r(\varphi) & = \frac{r_{\rm S}}{u(\varphi)} = \frac{r_{\rm S}}{x_1} = - r_{\rm S} \frac{\sigma_2(\vec \varphi_{\Theta})}{\sigma_1(\vec \varphi_{\Theta})}
\end{align} 
where $\vec \varphi_{\Theta} = \vek{\varphi_1}{\varphi-\varphi'_0}$. Here $\varphi'_0$ is given by
\begin{equation}
\varphi'_0 = \varphi_0 + \int_{u_0}^{\infty} dz_2 
\end{equation}
and $\varphi_1$ is chosen in such a way that $(2\omega)^{-1} \vec \varphi_{\Theta}$ is an element of the theta--divisor $\Theta_{K_\infty}$.

\paragraph{The classification of the solutions}

Having an analytical solution at hand we can explore the set of all possible solutions in a systematic manner. The shape of an orbit depends on 
$E$, $L$, and $\Lambda$ or, equivalently, on $\lambda$, $\mu$, and $\rho$
($r_{\rm S}$ can been absorbed through a rescaling of the radial coordinate). Since $r$ should be real and positive it is clear from \eqref{potential} that physically acceptable regions are given by those $u$ for which $E > V_{\rm eff}$. The zeros of $P_5$ are related to the points of intersection of $E$ and $V_{\rm eff}$, and a real and positive $P_5$ is equivalent to $E > V_{\rm eff}$. Hence, the number of positive real zeros of $P_5$ characterizes the form of the resulting orbit. A detailed discussion is presented in \cite{HackmannLaemmerzahl07}. 

If we denote by $e_1,\ldots,e_n$ the positive real zeros, then it follows that the physically acceptable regions are given by $[0,e_1], [e_2,e_3], \ldots, [e_n,\infty]$ if $n$ is even and by $[e_1,e_2], \ldots, [e_n,\infty]$ if $n$ is odd. With respect to $r$ we have the following classes of orbits:
\begin{enumerate}[(i)]\itemsep=-2pt
\item the region $[0,e_1]$ corresponds to escape orbits, 
\item the region $[e_n,\infty]$ corresponds to orbits falling into the singularity, i.e. to terminating orbits, and 
\item the regions $[e_i,e_{i+1}]$ correspond to bound orbits. 
\end{enumerate}
The case of no positive real zero corresponds to a particle coming from infinity and falling to the singularity. 

\begin{figure}[t]
\subfigure[][$\mu=0.92$, $\lambda=0.28$, $\Lambda =10^{-5} \text{km}^{-2}$, $r_0=133.60 \text{km}$]{
\includegraphics[width=0.22\textwidth]{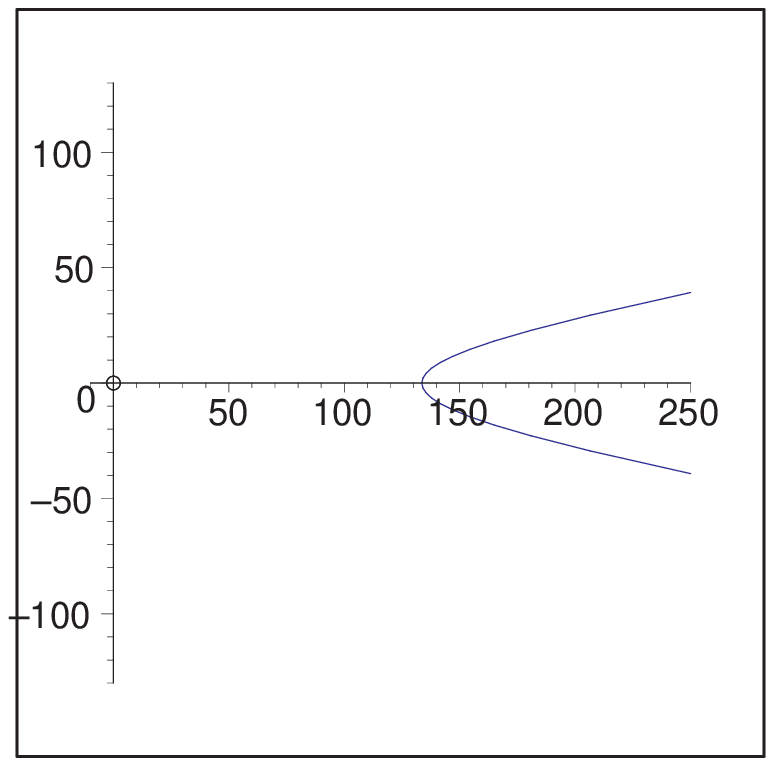}
}
\subfigure[][$\mu=1.1$, $\lambda=0.2$, $\Lambda = -10^{-5} \text{km}^{-2}$, $r_0=185.37 \text{km}$]{
\includegraphics[width=0.22\textwidth]{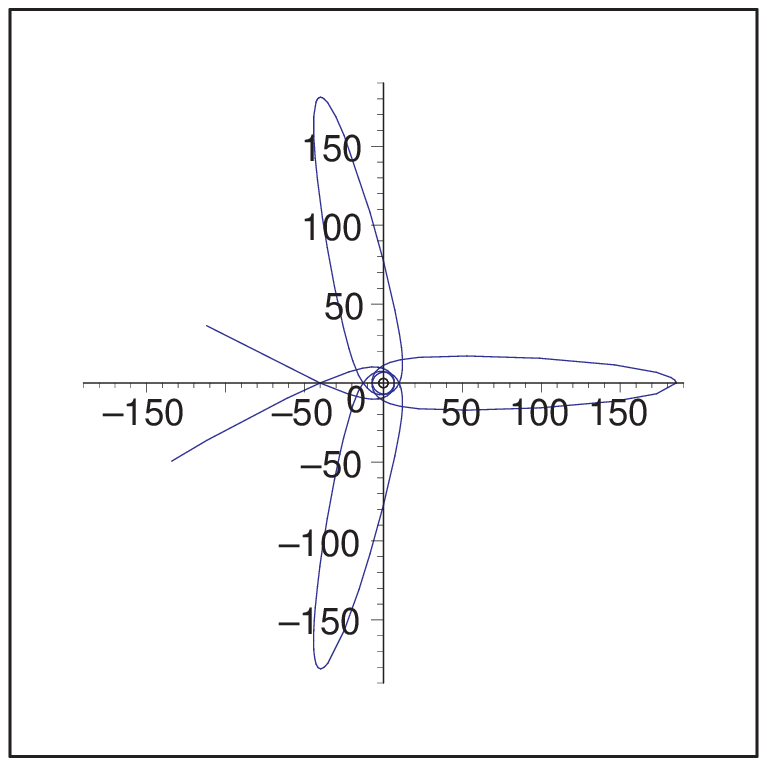}
}
\caption{Orbits with no analogue for $\Lambda=0$. (a): Reflection at the $\Lambda$--barrier (b): bound orbit with perihelion shift which would be an escape orbit for $\Lambda=0$. The scale is in km. \label{Fig:Orbits}}
\end{figure}

The number of real zeros are shown in Fig.~\ref{Dia}. A non--vanishing cosmological constant dramatically changes the possible structure of orbits. While for negative cosmological constant there are only bound orbits or orbits terminating in the singularity, for vanishing $\Lambda$ we have terminating, bound as well as escape orbits. For $\Lambda > 0$ we have in addition orbits which will be ``reflected'' at the potential barrier related to positive $\Lambda$. In Fig.~\ref{Fig:Orbits} two orbits without analogue for $\Lambda=0$ are shown. 

For bound orbits it is possible to directly identify the perihelion shift $\Delta_{\text{perihel}}$ by the difference of the periodicity of the solution $r(\varphi)$ from $2 \pi$. Let us assume that the bound orbit corresponds to the interval $[e_k,e_{k+1}]$ and that the path $a_i$ surrounds this interval. Then
\begin{equation}
\Delta_{\text{perihel}} = 2\pi - 2\omega_{2i} =  2 \int_{e_k}^{e_{k+1}} \frac{x dx}{\sqrt{P_5(x)}}.
\end{equation}

\paragraph{Summary and outlook}

We derived the complete set of solutions for the geodesic equation in a Schwarzschild--(anti)de Sitter space--time. A more detailed discussion of the approach will be published elsewhere \cite{HackmannLaemmerzahl07}. Beside the importance of such solutions mentioned in the introduction this result may also be applied to the study of the Pioneer anomaly \cite{HackmannLaemmerzahl07} which has been speculated to be due to the influence of the cosmological expansion. Although the cosmological constant is very small, its influence on nearly critical orbits is much stronger. As the anomaly is also small, a definite quantitative answer to the question whether the cosmological constant has an influence on the Pioneer orbits requires an analytical solution. Our result may also be applied to the motion of stars in the outer regions of galaxies. Furthermore, the above techniques apply to the geodesic equations in higher dimensional Schwarzschild, Schwarzschild-(anti)de Sitter, Reissner-Nordstr\"om and Reissner-Nordstr\"om-(anti)de Sitter space-times  \cite{HackmannKagramanovaetalpre}. The use of this method in more general space--times, namely the Plebanski--Demianski class without acceleration is under way \cite{HackmannKagramanovaetalpre}. In these cases polynomials of 6th order appear which can be treated with the above method.

We note that our method is not capable to solve equations of motions based on a polynomial of 7th or higher order which appear, e.g., in the geodesic equation in Schwarzschild space--times in higher than 7 dimensions. This will entail an Abel map with a preimage of dimension three or higher, which could only be reduced by one dimension with the presented method. 

\begin{acknowledgements}
We would like to thank H. Dittus, V. Kagramanova, J. Kunz, O. Lechtenfeld, D. Lorek and in particular P.H. Richter for fruitful discussions. Financial support of the German Aerospace Center DLR and of the German Research Foundation DFG is gratefully acknowledged. 
\end{acknowledgements}


\end{document}